\documentclass[aps,prl,twocolumn,showpacs]{revtex4}
\usepackage{amsfonts}
\usepackage{epsfig}
\usepackage{amsmath,amssymb}

\begin{document}
\title{Berry-Phase induced Heat Pumping and its Impact on the Fluctuation Theorem}

\author{Jie Ren$^{1,2}$}\email{renjie@nus.edu.sg}
\author{Peter H\"anggi$^{2,3}$}\email{hanggi@physik.uni-augsburg.de}
\author{Baowen Li$^{1,2}$}\email{phylibw@nus.edu.sg}

\affiliation{$^1$ NUS Graduate School for Integrative Sciences and
Engineering, Singapore 117456, Republic of Singapore \\ $^2$
Department of Physics and Centre for Computational Science and
Engineering, National University of Singapore, Singapore 117546,
Republic of Singapore \\ $^3$ Institut f\"ur Physik, University
Augsburg, Universit\"atsstr. 1, D-86135 Augsburg, Germany}

\date{\today}
\begin{abstract}
Applying adiabatic, cyclic two parameter modulations we
investigate quantum heat transfer across an anharmonic molecular
junction contacted with two heat baths. We demonstrate that the
pumped heat typically exhibits a Berry phase effect in providing
an additional geometric contribution to heat flux. Remarkably, a
robust fractional quantized geometric phonon response is
identified as well. The presence of this geometric phase
contribution in turn causes a breakdown of the fluctuation theorem
of the Gallavotti-Cohen type for quantum heat transfer. This can
be restored only if (i) the geometric phase contribution vanishes
and if (ii) the cyclic protocol preserves the detailed balance
symmetry.

\end{abstract}

\maketitle

Understanding and controlling of heat transfer due to phonons
occurring in low dimensional nanoscale systems is both of prime
and practical importance \cite{WL08}. Pioneering experimental
works carried out recently, such as nanotube thermal rectifier
\cite{exp_diode}, nanotube phonon waveguide \cite{waveguide} has
spawn {\it phononics}, i.e. the science and engineering of phonons
\cite{WL08}, as an emerging new scientific discipline where heat
flow can be manipulated as flexibly as electronic current.
Although the nonlinear (anharmonic) interaction has been
demonstrated as a crucial component \cite{Li06,Wu09} in various
functional thermal devices, the heat control has heretofore
typically been achieved by applying a temperature bias, for which
in accordance with the second law of thermodynamics -- heat flows
from ``hot" to ``cold" spontaneously.

Recent studies show that spontaneous, rare fluctuations of
anomalous heat transfer may occur \cite{RMP}, thus being seemingly
in apparent violation with the second law. Clearly, however, no
violation of the second law occurs on average. The typical measure
of such violations is the (small) probability for such anomalous
events as they emerge from a heat exchange fluctuation theorem
(FT) \cite{RMP, Jarzynski04, Saito07, Talkner09}. The FT for
(nonequilibrium) entropy production \cite{Evans1993,GC1995} and
heat flux \cite{Jarzynski04, Saito07} describes that the
distribution, $P_{\tau}(Q)$, of the heat $Q$ transferred from the
left ($L$) bath at temperature $T_L$ to the right ($R$) bath at
$T_R$ over a long time interval $\tau$, obeys the relation:
$\lim\limits_{\tau\rightarrow\infty}{\tau}^{-1}\ln[{P_{\tau}(Q)}/{P_{\tau}(-Q)}]=Q(\beta_R-\beta_L)/\tau,$
where $\beta_{L, R}=1/k_BT_{L, R}$. This FT thus shows explicitly
that  heat can transfer spontaneously from ``cold" to ``hot" with
finite, although typically with very small probability. In
particular, Ref. \cite{Saito07} demonstrates this FT in the
quantum case for heat transfer across a quantum harmonic chain
coupled with thermal reservoirs. A particular challenge that
arises is then whether this quantum Gallavotti-Cohen type FT
remains valid also in the nonlinear quantum regime beyond the
quantum harmonic chain limit, and, more generally, whether such a
heat-flux FT still can be formulated in presence of cyclic
time-dependent manipulations of certain control parameters.

\begin{figure}
\hspace{1mm} 
\scalebox{0.35}[0.32]{\includegraphics{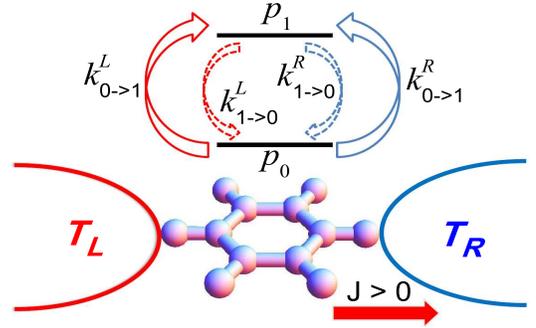}} \vspace{-3mm}
\caption{(Color online) A schematic representation of the
anharmonic molecular junction. Quantum heat transfer is generated
via a dynamics of excitation and relaxation of the local single
mode. The heat flux $J$ from the center to the right bath is
defined as positive.} \label{Fig0}
\end{figure}

In the context of time-dependent manipulations various molecular
heat pumps have been proposed to efficiently control heat flux
against thermal gradients at the nanoscale. In all those cases the
system is driven far away from equilibrium by use of an external
modulation imposed on system parameters. For example, a molecular
model with modulated energy levels, has been found to operate as a
heat pump \cite{Segal0608}. Likewise, a spin system leading to the
heat pumping has been studied with Ref. \cite{Dhar07}. Other
schemes investigated pumping of heat in electronic nanoscale
devices by applying time-periodic laser fields \cite{REY07}.
Moreover, Brownian heat motors  fueled by oscillating temperatures
have recently been devised as well \cite{Li0809,Ren09}. Given such
time-dependent manipulations one may therefore scrutinize whether
the physics of a non-vanishing geometric  phase does impact the
transfer of heat under external modulations. If so, what is its
impact on the existence of a heat-flux FT?

In this Letter, we shall answer these above mentioned objectives
by studying quantum heat transport across an anharmonic molecular
junction model. We start  with a system consisting of a molecular
junction coupled to two thermal baths \cite{Segal0608}, as
illustrated in Fig. \ref{Fig0}. The total Hamiltonian $H_{tot}$ is
composed of the following contributions:
$H_{tot}=H_{S}+H^L_B+H^R_B+V^L_{SB}+V^R_{SB}$: system Hamiltonian
$H_{S}=\sum_{n=0}^{N-1}E_n|n\rangle\langle n|$, with
$E_n=n\hbar\omega_0,$ where we assume that heat transport is
dominated by a single mode and thus consider a two-level system
($N=2$) to simulate the strong nonlinearity \cite{mastereq}. If
$N\rightarrow\infty$, the system reduces to the quantum harmonic
case. The two thermal baths are represented by sets of independent
harmonic modes, i.e.,
$H^{\nu}_B=\sum_{k}\hbar\omega_{k}b_{k,\nu}^{\dagger}b_{k,\nu}$,
with $\nu=L,R$, where $b_{k,\nu}^{\dagger}, b_{k,\nu}$ are the
bosonic creation and annihilation operators associated with the
phonon mode $k$ of bath $\nu$. The system-bath interactions is
taken to be bilinear, i.e.,  $V^{\nu}_{SB}=
B_{\nu}\sum_{n=1}^{N-1}\sqrt{n}|n\rangle\langle n-1| +
\mathrm{c.c.}$,
$B_{\nu}=\sum_{k}\gamma_{k,\nu}(b^{\dagger}_{k,\nu}+b_{k,\nu})$,
where the system-bath interaction is characterized by the phonon
spectral function $\Gamma_{\nu}(\omega) =2\pi\sum_k
\gamma_{k,\nu}^2 \delta(\omega-\omega_k)$. In the following, we
use wide-band limit $\Gamma_{\nu}(\omega)=\Gamma_{\nu}$. As shown
with Ref. \cite{mastereq}, in the limit of fast dephasing and
using the Redfield approximation for weak system-bath coupling,
the underlying dynamics can be modeled as follows:
\begin{eqnarray}\label{eq:master}
\dot{p}_1(t)=-p_1(t)(k^L_{1 \rightarrow 0}+k^R_{1\rightarrow
0})+p_0(t)(k^L_{0 \rightarrow 1}+k^R_{0\rightarrow 1}).
\end{eqnarray}
Here, $p_n(n=0,1)$ denotes the probability of the molecule to
occupy the state $|n\rangle$, satisfying $p_0(t)+p_1(t)=1$. The
activation and relaxation rates read:
\begin{eqnarray}\label{eq:detailed}
k^{\nu}_{0\rightarrow 1}=\Gamma_{\nu}N_{\nu}(\omega_0), \quad
k^{\nu}_{1\rightarrow 0}=\Gamma_{\nu}[N_{\nu}(\omega_0)+1],
\end{eqnarray}
where $N_{\nu}(\omega_0)=[e^{\beta_{\nu}\hbar\omega_0}-1]^{-1}$ is
the Bose-Einstein occupation probability. Finally, the
steady-state heat flux at the right contact (being equal to the
heat flux at the left contact) is expressed as
\begin{eqnarray}\label{eq:flux}
J=\hbar\omega_0[p_1^sk^R_{1\rightarrow 0}-p_0^sk^R_{0\rightarrow
1}],
\end{eqnarray}
where the superscript $s$ means the steady state. The first term
denotes the energy flux going from  the molecule into the bath $R$
while the second term provides the opposite heat flux from the
bath $R$ back into the system.

{\it Geometric Berry Phase induced Heat Pumping.} For heat pump
operation, the molecular junction connected to the two reservoirs
is subjected to cyclic parameter modulations. This could be
realized by imposing a modulation on either of the following
parameters: $\omega_0(t)$, $\Gamma_L(t)$, $\Gamma_R(t)$, $T_L(t)$,
$T_R(t)$. Throughout the following, the modulations acting on such
system parameters are assumed to be slow, i.e., we employ
adiabatic modulations. Let the period of modulation be
$\mathcal{T}_{p}=2\pi/\Omega$. The typical frequency for a
carbon-carbon bond is $1.4\times10^{14} s^{-1}$ \cite{carbon}.
$\Gamma_{\nu}$ is around $10^{15} s^{-1}$, according to the
measurement with alkane molecular junction \cite{Alkane}. The
relaxation time for fast thermalization usually is on the order of
a few fs or ps. Thus, the modulation time scale must obey
$2\pi/\Omega\gg1$ ps. In this way, the assumption of adiabatic
modulation is valid whenever the driving frequency $\Omega\ll 1$
THz.

Of prime interest is the heat flux from the molecule into the bath
$R$ during the long time span $\tau$. This is achieved upon
introducing the characteristic function for the phonon counting
field $\chi$, i.e., \cite{Gopich06,Sinitsyn07a}
\begin{eqnarray}\label{eq:CF}
\mathcal{Z}_{\tau}(\chi)=\sum_{q=-\infty}^{\infty}P_{\tau}(q)e^{iq\chi}=\mathbf{1}^{\dag}\hat{T}
\left[e^{-\int^{\tau}_{0}\mathcal{H}(\chi,t)dt}\right]\mathbf{p}(0),  \\
\mathcal{H}(\chi,t)\doteq \left[
\begin{array}{cc}
  k^L_{0\rightarrow1}+k^R_{0\rightarrow1} & -k^L_{1\rightarrow0}-k^R_{1\rightarrow0}e^{i\chi} \\
  -k^L_{0\rightarrow1}-k^R_{0\rightarrow1}e^{-i\chi} & k^L_{1\rightarrow0}+k^R_{1\rightarrow0} \\
\end{array}
\right],\label{eq:H}
\end{eqnarray}
where $P_{\tau}(q)$ is the probability distribution of having heat
$Q=q\hbar\omega_0$ transferred from the molecule into the bath $R$
during  time $\tau \rightarrow \infty$. Here,  $1^{\dag}=[1,1]$,
$\hat{T}$ denotes the time-ordering operator, and
$\mathbf{p}(0)=[p_0(0),p_1(0)]^T$ are the intial occupation
probabilities. Then, the cumulant generating function is obtained
as:
$\mathcal{G}(\chi)\equiv\lim\limits_{\tau\rightarrow\infty}\tau^{-1}\ln\mathcal{Z}_{\tau}(\chi),$
which generates 
the heat current via the relation:
$J=\hbar\omega_0\left.{\partial\mathcal{G}(\chi)}/{\partial(i\chi)}\right|_{\chi=0}$.
Denote by $\lambda_0(\chi,t)$ the instantaneous eigenvalue of
$\mathcal{H}(\chi,t)$ with the smallest real part and
$|\psi_0(\chi,t)\rangle$ ($\langle\varphi_0(\chi,t)|$) the
corresponding normalized right (left) eigenvector.  The cumulant
generating function takes on the following form, being composed of
two parts \cite{Sinitsyn07a,supply}, namely:
\begin{eqnarray}
\mathcal{Z}_{\tau}(\chi)&=&e^{\tau
\mathcal{G}}=e^{\tau(\mathcal{G}_{\mathrm{dyn}}+\mathcal{G}_{\mathrm{geom}})},\\
\mathcal{G}_{\mathrm{dyn}}&=&-\mathcal{T}_p^{-1}\int^{\mathcal{T}_p}_{0}dt
\lambda_0(\chi,t), \label{eq:Gcl} \\
\mathcal{G}_{\mathrm{geom}}&=&-\mathcal{T}_p^{-1}\int^{\mathcal{T}_p}_{0}dt
\langle\varphi_0|\partial_t|\psi_0\rangle, \label{eq:Ggeom}
\end{eqnarray}
The fist contribution $\mathcal{G}_{\mathrm{dyn}}$ presents the
temporal average and defines the dynamic heat transfer. This is
the only term which survives in the static limit.  The second,
geometric  part $\mathcal{G}_{\mathrm{geom}}$ presents an
additional contribution caused by the adiabatic cyclic evolution.
As we shall see it is this part which possesses  a nontrivial
geometric interpretation. Let us rewrite
$\mathcal{G}_{\mathrm{geom}}$ as a line integral over the closed
contour $\mathcal{R}$ in the parameter space $\mathbf{u}$:
\begin{eqnarray}
\mathcal{G}_{\mathrm{geom}}=
-\mathcal{T}_p^{-1}\oint_{\mathcal{R}}d\mathbf{u}\cdot\mathcal{A}_{\mathbf{u}},
\;\:\:
\mathcal{A}_{\mathbf{u}}=\langle\varphi_0(\mathbf{u})|\frac{\partial}{\partial_{\mathbf{u}}}|\psi_0(\mathbf{u})\rangle.
\end{eqnarray}
Thus, this is an analog of a Berry phase \cite{Berryphase}, which
does not contain time $t$ explicitly and only depends on the
geometry of the modulation contour in the parameter space
$\mathbf{u}$. In the case of two parameters being modulated, say
$u_1, u_2$, using Stokes theorem, we find
\begin{eqnarray}\label{eq:Gc}
\mathcal{G}_{\mathrm{geom}}=
-\mathcal{T}_p^{-1}\iint_{\mathcal{S_R}}du_1du_2\mathcal{F}_{u_1u_2},
\end{eqnarray}
where $\mathcal{S_R}$ is the integral area enclosed by the contour
$\mathcal{R}$.
\begin{eqnarray}\label{eq:curvature}
\mathcal{F}_{u_1u_2}=\langle\partial_{u_1}\varphi_0|\partial_{u_2}\psi_0\rangle
-\langle\partial_{u_2}\varphi_0|\partial_{u_1}\psi_0\rangle
\end{eqnarray}
is an analog of the gauge invariant Berry curvature
\cite{Berryphase}.

\begin{figure}
\hspace{1mm} {\hbox{\epsfxsize=81mm \epsffile{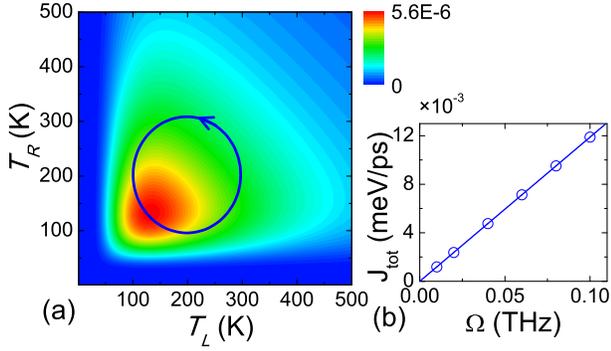}}}
\vspace{-3mm} \caption{(Color online) (a) The contour map of
$-\partial \mathcal{F}_{T_LT_R}(\chi)/\partial(i\chi)|_{\chi=0}$,
for $\Gamma_L=\Gamma_R$ and $\hbar\omega_0=25$ meV. The (blue)
circle with an arrow denotes the path of two-parameter temperature
modulations: $T_L(t)=200+100\cos(\Omega t+\pi/4)$,
$T_R(t)=200+100\sin(\Omega t+\pi/4)$. The integral area
$\mathcal{S_R}$ is within the circle. (b) Pure Berry-phase induced
heat current: $J_{\mathrm{tot}}=J_{\mathrm{geom}}$
($J_{\mathrm{dyn}}=0$). The straight line is the analytical result
from Eq.~(\ref{eq:Jgeom}), while the open circles give  the
simulation results by integrating Eq.~(\ref{eq:master}).}
\label{Fig2}
\end{figure}

Let us next specify the case that the bath temperatures $T_L(t),
T_R(t)$ are subjected to adiabatic modulations. Then Eq.
(\ref{eq:curvature}) yields the Berry curvature in  temperature
space, reading:
\begin{eqnarray}
\mathcal{F}_{T_LT_R}(\chi)=-
C_LC_R\frac{2i\sin(\chi)\Gamma_L\Gamma_R(\Gamma_L+\Gamma_R)}{(\sqrt{K^2+4D})^3},
\end{eqnarray}
where
$C_{\nu}=k_B\beta^2_{\nu}\hbar\omega_0e^{\beta_{\nu}\hbar\omega_0}N^2_{\nu}$,
$K=\Gamma_L(1+2N_L)+\Gamma_R(1+2N_R)$,
$D=\Gamma_L\Gamma_RN_LN_R(e^{\beta_R\hbar\omega_0}e_{+\chi}
+e^{\beta_L\hbar\omega_0}e_{-\chi})$ with $e_{\pm\chi}\equiv
e^{\pm i\chi}-1$. Upon substituting this Berry curvature into
Eq.~(\ref{eq:Gc}), the total heat flux emerges as:
\begin{eqnarray}
&&J_{\mathrm{tot}}=\hbar\omega_0\left.\frac{\partial[\mathcal{G}_{\mathrm{dyn}}(\chi)
+\mathcal{G}_{\mathrm{geom}}(\chi)]}{\partial
(i\chi)}\right|_{\chi=0}= J_{\mathrm{dyn}}+J_{\mathrm{geom}},
\nonumber \\
&&J_{\mathrm{dyn}}=\frac{\hbar\omega_0}{\mathcal{T}_p}\int^{\mathcal{T}_p}_0dt\frac{\Gamma_L\Gamma_R(N_L-N_R)}{K},
\\ \label{eq:Jgeom}
&&J_{\mathrm{geom}}=\frac{\hbar\omega_0}{\mathcal{T}_p}\iint_{\mathcal{S_R}}dT_LdT_R\left.\frac{-\partial
\mathcal{F}_{T_LT_R}(\chi)}{\partial(i\chi)}\right|_{\chi=0},
\end{eqnarray}
where
\begin{eqnarray}
\left.-\frac{\partial
\mathcal{F}_{T_LT_R}(\chi)}{\partial(i\chi)}\right|_{\chi=0}=\frac{2C_LC_R\Gamma_L\Gamma_R(\Gamma_L+\Gamma_R)}{K^3}.
\end{eqnarray}
The dynamic part $J_{\mathrm{dyn}}$ just coincides with the
temporal average of the heat flux obtained from $J \equiv J(t)$ in
Eq.~(\ref{eq:flux}). The geometric part $J_{\mathrm{geom}}$ is the
additional heat flux that results from the nontrivial Berry phase
effect. The ratio of this geometric heat flux and the dynamic one
is typically about $\Omega/\Gamma_{\nu}$. To avoid that
$J_{\mathrm{geom}}$ is masked by $J_{\mathrm{dyn}}$, we choose a
symmetric molecular junction with $\Gamma_L=\Gamma_R$, and
modulate $T_L(t), T_R(t)$ as indicated by the circle contour in
Fig.~\ref{Fig2}a. Then one finds that $J_{\mathrm{dyn}}\equiv0$
and $J_{\mathrm{geom}}\neq0$, see  Fig.~\ref{Fig2}b,  so that the
Berry phase induced $J_{\mathrm{geom}}$ dominates the heat
transport. This is the case for which the geometric phase effect
on heat transport is distinctly experimentally detectable. As a
main finding we have that the Berry phase effect acts as a heat
pump, providing  an additional heat flux across the molecular
junction even though on average no thermal bias acts and the
system is symmetric. Note also, distinct from the irreversible
heat flux $J_{\mathrm{dyn}}$, $J_{\mathrm{geom}}$ is
time-reversible, i.e., under the time-reversed modulation
($t\rightarrow -t$) the Berry-phase induced heat flux just
reverses sign.

{\it Fractional Quantization of Phonon Response.} Remarkably, we
find a fractional quantized phonon response for large temperature
driving: the integral in Eq.~(\ref{eq:Jgeom}) can be rewritten as
$\int^{\infty}_{0}\int^{\infty}_{0} dN_LdN_R
2\Gamma_L\Gamma_R(\Gamma_L+\Gamma_R)/K^3=1/4$, yielding
\cite{supply}:
\begin{eqnarray}
J_{\mathrm{geom}}=\frac{1}{4}\hbar\omega_0/\mathcal{T}_p.
\end{eqnarray}
This $1/4$ fractional quantized geometric phonon response is
robust since it does not depend on the specific values of
$\hbar\omega_0, \Gamma_L, \Gamma_R$. It means that the geometric
phase effect caused by two bath temperature modulations is able to
pump maximally on average one phonon $\hbar\omega_0$ per four
cycles.

{\it Impact on Heat-Flux Fluctuation Theorem.} Besides the dynamic
$\mathcal{G}_{\mathrm{dyn}}$, $\mathcal{G}_{\mathrm{geom}}$ will
generally not only contribute additionally to the average heat
transfer but also impact the higher moments of the heat current
(such as the phonon counting statistics) and other heat transport
characteristics as well. In the following, we study its impact on
the heat-flux fluctuation theorem. Before doing so, let us address
first the static situation with
$\mathcal{G}_{\mathrm{geom}}(\chi)\equiv0$, yielding
\begin{eqnarray}\label{eq:G}
\mathcal{G}(\chi)=
\mathcal{G}_{\mathrm{dyn}}(\chi)=-\lambda_0(\chi)=\frac{-K+\sqrt{K^2+4D}}{2}.
\end{eqnarray}
Then, $\lambda_0$ obeys the Gallavotti-Cohen (GC) symmetry
\cite{Lebowitz99} (and alike for
$\mathcal{G}_{\mathrm{dyn}}(\chi)$, $\mathcal{G}(\chi)$ and
$\mathcal{Z}_{\tau}(\chi)$), reading
\begin{eqnarray} \label{eq:symm}
\mathcal{\lambda}_0(\chi)=\mathcal{\lambda}_0(-\chi+i\beta^*),
\end{eqnarray}
where $\beta^*=\ln[(k^L_{0\rightarrow1}k^R_{1\rightarrow0})/
(k^R_{0\rightarrow1}k^L_{1\rightarrow0})]$. In virtue of
Eq.~(\ref{eq:detailed}), yielding the  detailed balance relation
$k^{\nu}_{0\rightarrow1}=k^{\nu}_{1\rightarrow0}e^{-\beta_{\nu}\hbar\omega_0},$
we find that $\beta^*=\hbar\omega_0(\beta_R-\beta_L)$. Via an
inverse Fourier transform of Eq.~(\ref{eq:CF}), this  GC symmetry
results in the {\it quantum} FT of heat transport for an {\it
anharmonic} molecular junctions, reading with $Q=q\hbar\omega_0$:
\begin{eqnarray} \label{eq:FT}
\lim\limits_{\tau\rightarrow\infty}\frac{1}{\tau}\ln[\frac{P_{\tau}(Q)}{P_{\tau}(-Q)}]=Q(\beta_R-\beta_L)/\tau,
\end{eqnarray}
which precisely coincides (without any correction) with the result
for the quantum harmonic chain \cite{Saito07}. This FT gives the
probability of observing spontaneous  ``second law violation":
Assume $T_L<T_R$, i.e. $\beta^*<0$; the upper bound to observe the
``violation" for spontaneous  heat transfer from (left) cool  to
(right) hot is  estimated as: $\int^{\infty}_{c}dq
P_{\tau}(q)=\int^{\infty}_{c}dq P_{\tau}(-q) e^{q\beta^*}\leq
e^{c\beta^*}.$ It indicates that in absence of external
modulations, the probability of  at least $c$ phonons (or net
energy $c\hbar\omega_0$) transporting against the thermal bias is
nonvanishing detectable, although decaying exponentially.

For the  time-modulated  system the GC symmetry  ceases to hold
when $\mathcal{G}_{\mathrm{geom}}(\chi)\neq 0$. For example, in
the case of cyclic temperature modulations $T_L(t)$ and $T_R(t)$,
the Berry curvature $\mathcal{F}_{T_LT_R}(\chi)$ contains the
factor $\sin(\chi)$, which explicitly breaks the GC symmetry of
$\mathcal{G}_{\mathrm{geom}}(\chi)$, and alike for
$\mathcal{G}(\chi)$ and $\mathcal{Z}_{\tau}(\chi)$. Thus, the FT
Eq.~(\ref{eq:FT}) becomes  {\it violated} as a consequence of a
geometric phase induced breakdown of  GC symmetry. Moreover, even
for parameter modulations yielding
$\mathcal{G}_{\mathrm{geom}}(\chi)=0$, and with time-dependent
$\beta^*\rightarrow\beta^*(t)$, the GC symmetry for
$\mathcal{G}_{\mathrm{dyn}}(\chi)=-\mathcal{T}^{-1}_p\int^{\mathcal{T}_p}_{0}dt
\lambda_0(\chi,t)$ generally cannot be recovered, despite
$\lambda_0(\chi,t)=\lambda_0(-\chi+i\beta^*(t),t)$.

Interestingly, we find that for time modulations of the
system-bath couplings $\Gamma_L(t), \Gamma_R(t)$, the detailed
balance relation
$k^{\nu}_{0\rightarrow1}/k^{\nu}_{1\rightarrow0}=e^{-\beta_{\nu}\hbar\omega_0}$
remains intact, thus providing a  vanishing Berry curvature
$\mathcal{F}_{\Gamma_L\Gamma_R}(\chi)\equiv0$. Meanwhile, with the
resulting time-independent $\beta^*(t)=\beta^*$, one finds that
the GC symmetry of
$\mathcal{G}_{\mathrm{dyn}}(\chi)=-\mathcal{T}_p^{-1}\int^{\mathcal{T}_p}_{0}dt
\lambda_0(\chi,t)$ still holds. Consequently, we obtain a
vanishing Berry-phase induced heat pumping and, surprisingly as
well, also no violation of the FT, no matter how $\Gamma_{L}(t)$
and $\Gamma_{R}(t)$ are modulated.

In summary, through investigating  heat transport across an
anharmonic molecular junction by applying  cyclic two-parameter
modulations, we find that the system generally undergoes, apart
from dynamic pumping, also  a Berry phase induced heat pumping.
This geometric contribution exhibits a  robust fractional
quantized phonon response. Furthermore, the quantum  FT for heat
transport in presence of a static temperature bias holds true in
the anharmonic case as well. The presence of the  geometric phase,
however, violates the heat-flux FT. Only in situations of
vanishing Berry curvature and restoration of detailed balance
symmetry can the validity of the FT be recovered.

Although our present work did focus on the adiabatic regime, it
likely can be extended to the case of a non-adiabatic geometric
phase \cite{Ohkubo}, and maybe also for non-cyclic modulation
schemes in the spirit of \cite{Sinitsyn07b}. Because the geometric
phase has profound effects on material properties
\cite{Berryphase} we hope that our present findings do invigorate
others to undertake related studies aimed at uncovering intriguing
novel geometric phase induced thermal effects (such as
thermoelectricity) which will enrich further the discipline of
{\it phononics}.

The work is supported by the grant R-144-000-222-646 from National
University of Singapore (NUS) (J.R. and B.L.) and by the German
Excellence Initiative via the Nanosystems Initiative Munich (NIM)
and DFG priority program SPP 1243 (P.H.).

\begin{widetext}

\section{Supplementary Information of ``Berry-Phase induced Heat
Pumping and its Impact on the Fluctuation Theorem"}

%
%
\subsection{The Generating Function with the Phonon Counting Field
and associated Geometric Phase} In this section, we derive the
geometric phase effect from time evolution of the generating
function with phonon counting fields in more detailed form. For
our system, $P_{t}(q)$ denotes the probability that within time
$t$ the net number of phonons transferred from the molecule into
the bath $R$ is $q$. We then split this probability into two
parts, namely, $P_{t}(q)\doteq P_{t}(0,q)+P_{t}(1,q)$, where
$P_{t}(0,q)$ $(P_{t}(1,q))$ denotes the probability that having
$q$ net phonons transferred from the molecule into the bath $R$,
within time $t$, while the molecule is dwelling on the low ``0"
(high ``1") energy level at time $t$. The time evolution of our
system can be described as:
\begin{eqnarray}\label{eq:ME}
\frac{d}{dt}P_{t}(0,q)&=&-(k^L_{0\rightarrow1}+k^R_{0\rightarrow1})P_{t}(0,q)
+k^L_{1\rightarrow0}P_{t}(1,q)+k^R_{1\rightarrow0}P_{t}(1,q-1), \nonumber\\
\frac{d}{dt}P_{t}(1,q)&=&
k^L_{0\rightarrow1}P_{t}(0,q)+k^R_{0\rightarrow1}P_{t}(0,q+1)-(k^L_{1\rightarrow0}+k^R_{1\rightarrow0})P_{t}(1,q).
\nonumber
\end{eqnarray}
By multiplying the factor $e^{iq\chi}$ on both sides of the
equations and summing over $q$, we obtain the time evolution
equation of these individual generating functions:
$$\frac{d}{dt}|\Psi(\chi,t)\rangle=-\mathcal{H}(\chi,t)|\Psi(\chi,t)\rangle, \eqno(\mathrm{S1})$$
where $$\mathcal{H}(\chi,t)=\left[
\begin{array}{cc}
  k^L_{0\rightarrow1}+k^R_{0\rightarrow1} & -k^L_{1\rightarrow0}-k^R_{1\rightarrow0}e^{i\chi} \\
  -k^L_{0\rightarrow1}-k^R_{0\rightarrow1}e^{-i\chi} & k^L_{1\rightarrow0}+k^R_{1\rightarrow0} \\
\end{array}
\right],\;\;\;\;\;\; |\Psi(\chi,t)\rangle=\left(\begin{array}{c}
  \sum^{\infty}_{q=-\infty}P_{t}(0,q)e^{iq\chi} \\
  \sum^{\infty}_{q=-\infty}P_{t}(1,q)e^{iq\chi} \\
\end{array}\right).$$
Therefore, the characteristic function $\mathcal{Z}_{\tau}(\chi)$
emerges as:
$$\mathcal{Z}_{\tau}(\chi)\equiv\sum^{\infty}_{q=-\infty}P_{\tau}(q)e^{iq\chi}=\mathbf{1}^{\dag}|\Psi(\chi,\tau)\rangle=\mathbf{1}^{\dag}\hat{T}\left[e^{-\int^{\tau}_{0}\mathcal{H}(\chi,t)dt}\right]|\Psi(\chi,0)\rangle. \eqno(\mathrm{S2})$$
At initial $t=0$, the net number of transferred phonons is
$0$, thus $|\Psi(\chi,0)\rangle=[p_0(t=0),p_1(t=0)]^T$. 
Note that the time evolution equation Eq. (S1) for generating
functions including the counting field exists for systems
described by such master equations generally\cite{Gopich2006}.

In the following, we are going to derive  the geometric phase from
the time evolution of the generating function (we omit the
counting field variable $\chi$ for the sake of better
readability). We stress that the time-dependent operator
$\mathcal{H}(t)$ is not a genuine, hermitian Hamilton operator.
However, we still can decompose it as
$$\mathcal{H}(t)|\psi_n(t)\rangle=\lambda_n(t)|\psi_n(t)\rangle,\;\;\;\;\;\; \langle\varphi_n(t)|\mathcal{H}(t)=\langle\varphi_n(t)|\lambda_n(t),$$
where $\lambda_n(t)$ denotes the instantaneous eigenvalue of
$\mathcal{H}(t)$ and $|\psi_n(t)\rangle$ $(\langle\varphi_n(t)|)$
the corresponding normalized right (left) eigenvector, satisfying
$\langle\varphi_m(t)|\psi_n(t)\rangle=\delta_{mn}$. Similar to
Berry's reasoning \cite{Berry1984}, we expand $|\Psi(t)\rangle$ as
$$|\Psi(t)\rangle=\sum_{n}a_n(t)\exp\left(-\int^{t}_{0}\lambda_n(t')dt'\right)|\psi_n(t)\rangle.
\eqno(\mathrm{S3})$$ By substituting Eq.~(S3) into the time
evolution equation Eq.~(S1), we find
$$\sum_{n}\dot{a}_n(t)\exp\left(-\int^{t}_{0}\lambda_n(t')dt'\right)|\psi_n(t)\rangle=\sum_{n}a_n(t)\exp\left(-\int^{t}_{0}\lambda_n(t')dt'\right)|\dot{\psi}_n(t)\rangle.
\eqno(\mathrm{S4})$$ Upon the multiplication of the left
eigenvector $\langle\varphi_m(t)|$ and observing that
 $\langle\varphi_m(t)|\psi_n(t)\rangle=\delta_{mn}$, we
obtain
$$\dot{a}_m(t)=-a_m(t)\langle\varphi_m(t)|\dot{\psi}_m(t)\rangle-\sum_{n\neq
m}a_n(t)\exp\left(-\int^{t}_{0}[\lambda_n(t')-\lambda_m(t')]dt'\right)\langle\varphi_m(t)|\dot{\psi}_n(t)\rangle.
\eqno(\mathrm{S5})$$ Note $\lambda_n$ are complex-valued
eigenvalues. The long time behavior of the system is governed by
the eigenmode $m=0$, whose eigenvalue $\lambda_0$ possesses the
smallest real part. Consequently, by neglecting the last term
within our adiabatic approximation, one obtains:
$$a_0(\tau)=\exp\left(-\int^{\tau}_{0}\langle\varphi_0|\dot{\psi}_0\rangle dt\right)a_0(0),\;\:\: \tau\rightarrow\infty. \eqno(\mathrm{S6})$$
Taking into account the adiabatic cyclic evolution over a
long-time period $\mathcal{T}_p$, we end up with
$$\mathcal{Z}_{\tau}=\mathbf{1}^{\dag}|\Psi(\tau)\rangle\approx\exp\left(-\frac{\tau}{\mathcal{T}_p}\int^{\mathcal{T}_p}_{0}dt\left[\lambda_0(t)+\langle\varphi_0|\dot{\psi}_0\rangle\right]\right)a_0(0)\mathbf{1}^{\dag}|\psi_0(\tau)\rangle, \eqno(\mathrm{S7})$$
which tells us that the cumulant generating function
$\mathcal{G}(\chi)\equiv\lim\limits_{\tau\rightarrow\infty}\tau^{-1}\ln\mathcal{Z}_{\tau}(\chi)=\mathcal{G}_{\mathrm{dyn}}+\mathcal{G}_{\mathrm{geom}}$
contains two contributions, one originating from the dynamic phase
factor and the other from the geometric phase factor (the boundary
value contribution:
$\tau^{-1}\ln\left[a_0(0)\mathbf{1}^{\dag}|\psi_0(\tau)\rangle\right]$
becomes negligible in the long time limit):
$$\mathcal{G}_{\mathrm{dyn}}=-\mathcal{T}_p^{-1}\int^{\mathcal{T}_p}_{0}dt
\lambda_0(\chi,t),\eqno(\mathrm{S8})$$
$$\mathcal{G}_{\mathrm{geom}}=-\mathcal{T}_p^{-1}\int^{\mathcal{T}_p}_{0}dt
\langle\varphi_0|\partial_t|\psi_0\rangle. \eqno(\mathrm{S9})$$
These two expressions of $\mathcal{G}_{\mathrm{dyn}}$ and
$\mathcal{G}_{\mathrm{geom}}$ coincide precisely with those
obtained in Ref. \cite{Sinitsyn2007a}, although derived therein
using a different approach.

\subsection{Fractional Quantized Phonon Response} Now, we give a
detailed explanation of the physical picture of the observed $1/4$
fractional quantized phonon response, in particular, the cause of
the Berry phase effect induced by temperature modulations of the
two heat baths to transfer on average $1/4$ phonon $\hbar\omega_0$
per one modulation cycle.

\begin{figure}
\hspace{1mm} 
\scalebox{0.55}[0.55]{\includegraphics{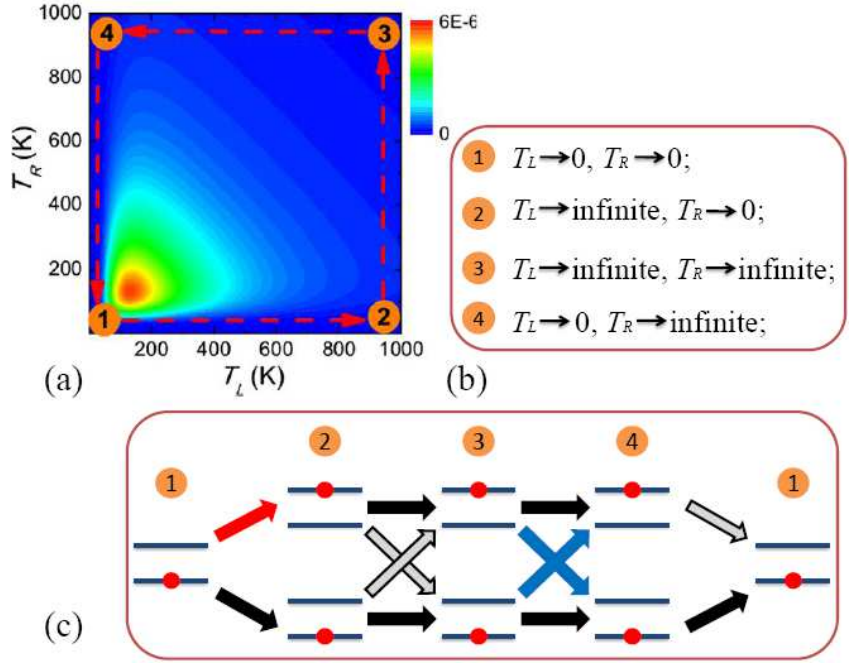}} \vspace{-3mm}
\caption{(Color online) A schematic representation of the
temperature modulation cycle and corresponding level transitions.
(a)The contour map of $-\partial
\mathcal{F}_{T_LT_R}(\chi)/\partial(i\chi)|_{\chi=0}$, for
$\Gamma_L=\Gamma_R$ and $\hbar\omega_0=25$ meV. The (red) dash
line with an arrow denotes the trajectory of temperature
modulations. (b)The temperature list of state (1)(2)(3)(4). (c)
The corresponding transitions during cycle
$(1)\rightarrow(2)\rightarrow(3)\rightarrow(4)\rightarrow(1)$.
The red arrow denotes only the up transition from bath $L$ is
excited; while the transitions only allowed at the $R$ bath are
indicated by blue arrows. Besides, black arrows indicate that the
system stays intact and gray arrows depict the transitions
occurring at both $L$ and $R$ bath with equal probability.}
\label{sFig1}
\end{figure}

Mathematically, when the integral area $\mathcal{S_R}$  due to the
two parameter modulation encloses a large part of the area around
the maximum of $-\partial
\mathcal{F}_{T_LT_R}(\chi)/\partial(i\chi)|_{\chi=0}$, or, put
differently for strong temperature driving temperature comprising
near zero temperature up to large values beyond  $\hbar\omega_0$,
as sketched by the dashed path in Fig.~\ref{sFig1}a, the integral
in $J_{\mathrm{geom}}$ can be recast as
$$J_{\mathrm{geom}}=\frac{\hbar\omega_0}{\mathcal{T}_p}\int^{\infty}_{0}\int^{\infty}_{0} dN_LdN_R
\frac{2\Gamma_L\Gamma_R(\Gamma_L+\Gamma_R)}{K^3}=\frac{1}{4}{\hbar\omega_0}/{\mathcal{T}_p},
\eqno(\mathrm{S10})$$

The underlying physical mechanism is as the following: As shown in
Fig.~\ref{sFig1}a, the trajectory of this so obtained temperature
modulation follows the path:
$(1)\rightarrow(2)\rightarrow(3)\rightarrow(4)\rightarrow(1)$,
where (1) ($T_L=0, T_R=0$), (2) ($T_L=0, T_R\rightarrow\infty$),
(3) ($T_L\rightarrow\infty, T_R\rightarrow\infty$), (4)
($T_L\rightarrow\infty, T_R=0$) as detailed in Fig.~\ref{sFig1}b.
For the two-level system under consideration, only for the
parameter set around (1) does the system  fully occupy the lower
level; at the other three sets (2), (3), (4), the system occupies
the upper level and lower one with equal probability. The level
transitions during the course of  the temperature modulation are
illustrated with Fig.~\ref{sFig1}c. Usually, the transition from
the lower (upper) level to the upper (lower) contain two parts of
contributions: one is from the left bath and the other is from the
right bath, respectively (see Fig. 1 in the text). However, for
the transition from (1) to (2), the temperature of bath $L$ is
increased from $0$ to $\infty$ so that only the up transition from
bath $L$ is excited, which is indicated as the red arrow; while
for the transition from (3) to (4), $T_L$ is modulated near zero
and $T_R$ keeps extremely high so that only the transitions from
the $R$ bath become allowed. Those are  indicated as blue arrows
(see Fig.~\ref{sFig1}c). Besides, black arrows indicate that the
system stays intact and gray arrows depict the transitions
occurring at either $L$ or $R$ bath with equal probability. Note
that only the up transition at $L$ followed by one down transition
at $R$ counts for the positive energy transport from $L$ to $R$.

The transitions during temperature modulations can be further
decomposed into eight individual paths, as illustrated in Fig.
\ref{sFig2}. Therefore, we can count the amount of energy
transport for each path individually. For example, in path (a),
the up transition is excited at $L$ during $(1)\rightarrow(2)$,
and then the system keeps in the upper level until the down
transition happens during $(4)\rightarrow(1)$. Since the down
transition exhibit a splitting, occurring at $L$ or $R$,  this
path contributes $1/2$ phonon for the energy transport. In path
(b), the up transition at $L$ is first excited and then during
$(3)\rightarrow(4)$ the down transition occurs at $R$, so that one
phonon is transported during the whole modulation cycle. The
amount of energy transport in other paths can be accounted
likewise.

Finally, the total amount of phonon transport during one
temperature modulation cycle is expressed as the average of
contributions from these eight routes:
$$\frac{1/2+1+0+1/2+0+(-1/2)+0+1/2}{8}=\frac{1}{4}. \eqno(\mathrm{S11})$$
In other words, after one complete modulation cycle, the system
returns to its ``original" state  but with $\hbar\omega_0/4$
energy given away. This effect has a geometric interpretation and
can be utilized to act as a heat pump.
\begin{figure}
\hspace{1mm} 
\scalebox{0.6}[0.6]{\includegraphics{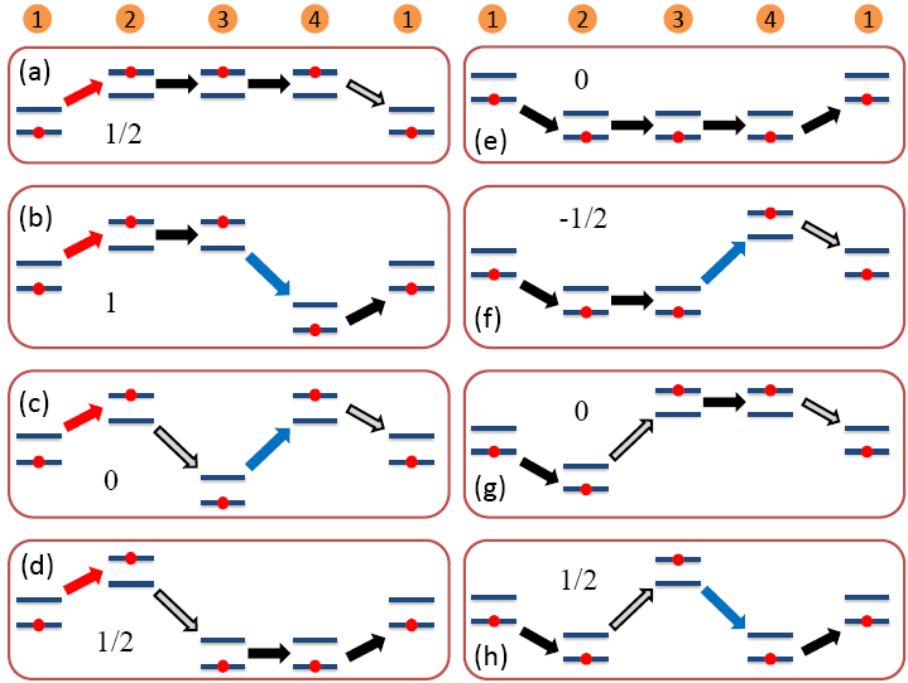}} \vspace{-3mm}
\caption{(Color online) Decomposition of the level transitions
into eight individual transition paths. The amount of phonon
transport contributed by each path is indicated.} \label{sFig2}
\end{figure}

In summary, this scenario makes it plausible that the Berry phase
effect caused by two-parameter temperature modulations is able to
induce on average a heat transfer across the molecular junction in
the amount of $1/4$ phonon $\hbar\omega_0$ per modulation cycle.
This $1/4$ fractional quantized phonon response is robust in the
sense that it does not depend on the specific values of
$\hbar\omega_0, \Gamma_L, \Gamma_R$.
\end{widetext}

\end{document}